\documentclass[]{raa}

\usepackage{graphicx, times}  
\usepackage{amsmath}

\newcommand{\teff}{$T_{\rm eff}$}
\newcommand{\logg}{log$g$}
\newcommand{\feh}{\rm [Fe/H]}
\newcommand{\MK}{${\rm M}_K$}

\begin{document}
\title{Red Clump Stars from the LAMOST data I: identification and distance}

  \volnopage{Vol.0 (200x) No.0, 000--000}      
   \setcounter{page}{1}          

\author{Junchen Wan\inst{1}, Chao Liu\inst{1}, Licai Deng\inst{1}, Wenyuan Cui\inst{2}, Yong Zhang\inst{3}, Yonghui Hou\inst{3}, Ming Yang\inst{1}, Yue Wu\inst{1}}

\institute{$^1$Key Laboratory of Optical Astronomy, National Astronomical Observatories, Chinese Academy of Sciences, 20A Datun Road, Beijing, 100012, China;  {junchenwan@bao.ac.cn}\\
$^2$Hebei Normal University, Shijiazhuang, China\\
$^3$Nanjing Institute of Astronomical Optics \& Technology, National Astronomical Observatories, Chinese Academy of Sciences, Nanjing 210042, China\\}

     \date{Received~~2009~~month day; accepted~~2009~~month day}

\abstract{
We present a sample of about 120,000 red clump candidates selected from the LAMOST DR2 catalog based on the empirical distribution model in the effective temperature vs. surface gravity plane. Although, in general, red clump stars are considered as the standard candle, they do not exactly stay in a narrow range of absolute magnitude, but may extend to more than 1 magnitude depending on their initial mass. Consequently, conventional oversimplified distance estimations with assumption of fixed luminosity may lead to systematic bias related to the initial mass or the age, which may potentially affect the study of the evolution of the Galaxy with red clump stars. We therefore employ an isochrone-based method to estimate the absolute magnitude of red clump stars from their observed surface gravities, effective temperatures, and metallicities. We verify that the estimation well removes the systematics and provide an initial mass/age independent distance estimates with accuracy less than 10\%.
}

\authorrunning{J. C. Wan et al.}            
   \titlerunning{Red clump stars from the LAMOST data I}  

   \maketitle

\section{Introduction}
Red clump (RC) stars are metal-rich stars in the evolution phase of helium-core burning (Cassisi \& Salaris~\cite{cassisi97}). They play important roles in the study of the Galactic disk because they are widespread in the thin disk, usually considered as the standard candle (Pacy\'nski \& Stanek ~\cite{pac98}) and luminous (Girardi et al.~\cite{gira98}; Alves ~\cite{alv00}; Groenewegen~\cite{gro08}) and prominent population in color magnitude diagram, which makes them be easily identified from multi-band photometry (L\'opez-Corredoira et al.~\cite{Lop02}).

Identification of individual field RC stars, however, is not trivial because they do highly overlap with the RGB stars. Pacy\'nski \& Stanek ~(\cite{pac98}) used a Gaussian to model the distribution of magnitude for RC stars and a quadratic polynomial for RGB stars. Then the stars located in the Gaussian dominated region are very likely to be RC stars. This method was applied by Nataf et al.(~\cite{nataf13}) to select the red clump stars in the Galactic bulge. Recently, Bovy et al.(~\cite{bovy14}) employed a new method to identify the RC stars based on their distribution in color--effective temperature--surface gravity--metallicity and stellar evolution model. However, the authors only identify the primary RC stars and remove the secondary population to simplify the distance estimation.

Compared to the identification, the distance estimation of RC stars are relatively simple with the assumption that the absolute magnitude of RC stars is around a fixed value with a small dispersion (Pacy\'nski \& Stanek ~\cite{pac98}; Girardi et al.~\cite{gira98}; Alves ~\cite{alv00}; Groenewegen~\cite{gro08}; Liu et al.~\cite{liu12}). 

However, the stellar evolution model demonstrates that the RC stars are not always stay in the same luminosity. They are separated into two subclasses: the primary RC stars, which have electronic-degenerate cores, and the second RC stars, which contain non-degenerate He-cores (Girardi~\cite{girardi99}). In general, the primary RC stars are low-mass and hence older, while the secondary RC stars are massive and therefore younger than 1\,Gyr. Most of the primary RC stars are generally brighter than the secondary RC stars, thus the former have smaller \logg\ than the later (see Stello et al.~\cite{stello13}). In the context of the evolution of the Galactic disk, we intend to obtain a sample of RC stars with a wide range of the age so that we can trace the evolution of the Galaxy from precent day back to $\sim10$\,Gyr. For instance, Salaris \& Girardi~(\cite{sal02}) fitted the distribution of the \emph{Hipparcos} (Perryman et al.~\cite{perryman97}) observed RC stars, including both the primary and secondary populations, in color-magnitude diagram with a stellar evolution model and derived the distribution of age in the solar neighborhood. Although keeping both the primary and secondary RC stars in the sample is important in the study of the disk evolution, the distance estimation may turn out to be more complicated since the RC stars would not be the standard candle any more in this case. 

As the first paper of a series of works on the evolution of the Galactic disk with the RC stars from the LAMOST catalog, we firstly develop a new method of identification for both the primary and secondary RC stars and a normal approach of distance estimation to both populations. This paper is organized as below. In Section~\ref{sect:identifyRC}, we briefly introduce the LAMOST survey data and describe the empirical method of identification of RC stars. In Section~\ref{sect:dist}, we develop the new method of the distance estimation. The performance of the estimates is then assessed in the same section. Finally, we further discuss the accuracy of our distance estimation in Section~\ref{sect:disc} and draw a brief conclusion in Section~\ref{sect:con}.

\section{Identification of RC stars}\label{sect:identifyRC}
In the section we identify the red clump stars from the LAMOST data using an empirical method.

\subsection{The LAMOST data}\label{sect:lamostdata}
The LAMOST, also called Guo Shou Jing telescope, is a 4-meter reflected Schmidt telescope with 4000 fibers on the 20-squared degree focal plane (Cui et al.~\cite{cui12}; Zhao et al.~\cite{zhao12}). The LAMOST survey will finally observe more than 5 million low resolution stellar spectra after its 5-year survey (Deng et al.~\cite{deng12}; Liu et al.~\cite{liu14a}). According to Yao et al.~(\cite{yao12}), the site of LAMOST prefers to observe the winter sky, which covers the Galactic anti-center region. Therefore, there will be lots of disk populations located in the region to be observed by the LAMOST. In this work. we adopt the derived \teff\ directly from the LAMOST pipeline (Wu et al.~\cite{wu11a}; Wu et al.~\cite{wu11b}; Wu et al.~\cite{wu14}; Luo et al. in prep) and the estimated \logg\ from Liu et al.~(\cite{liu15}), who estimate \logg\ using support vector regression with the training dataset from the Kepler asteroseismic \logg. The uncertainty of the \logg\ estimates is only about 0.1\,dex, which is a factor of 2 better than that from the LAMOST pipeline, for the metal-rich (\feh$>-0.6$) giant stars, including the RC stars. 

We select the stars with \logg\ between 0.9 and 3.5, \feh\ between -0.6 and 0.4, and \teff\ between 3600 and 6000\,K from the LAMOST DR2 catalog, which widely cover the RC region. We also remove all spectra with signal-to-noise ratio lower than 10 and finally obtain 279,423 stars. 

\subsection{Identification of RC stars}\label{sect:}
\begin{figure}
\centering
\includegraphics[scale=0.9]{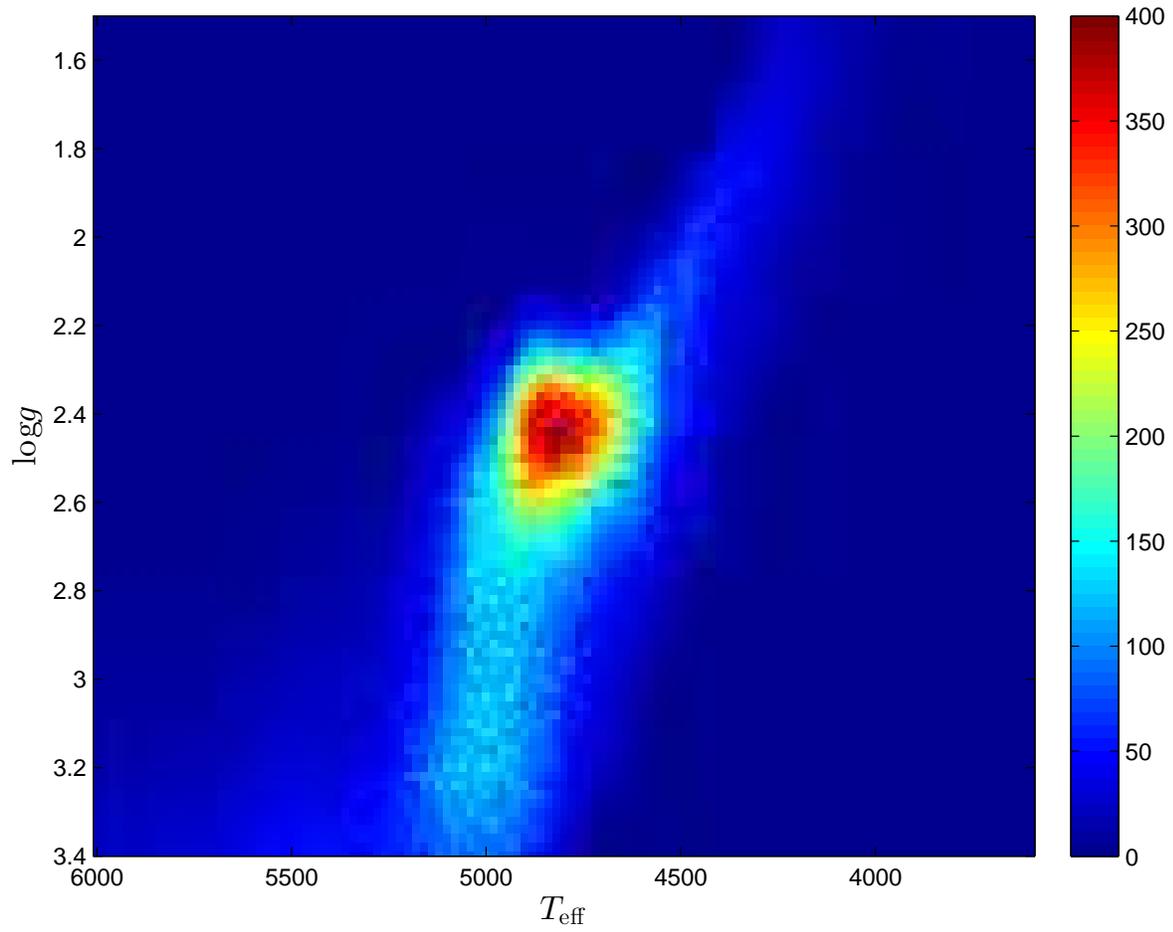}
\caption{The observed distribution of the metal-rich (\feh\,$>-0.6$) giant branch stars from LAMOST DR2 in log$g$ vs. $T_{\rm{eff}}$ plane. The bin size is $\Delta{\rm log}g = 0.02$ and $\Delta{T_{\rm eff}} = 20$K.}
\label{fig1}
\end{figure}

It is expected that lots of red clump stars in the disk population will be sampled by LAMOST survey. Indeed, Fig~\ref{fig1} shows that the RC stars from LAMOST DR2 catalog are prominent in \teff\ vs. \logg\ plane. In this section, we establish an approach of identification of RC stars from the LAMOST data.

Paczy\'{n}ski \& Stanek~(\cite{pac98}) used a Gaussian and quadratic polynomial to model the distribution of the magnitude for the RC and RGB stars, respectively. We expand this method into two dimensions in \teff\ vs. \logg\ plane.

First, we empirically build a 2-D distribution model for the RGB stars in \teff\ vs. \logg\ plane with various metallicity bins ($\feh\,=(-0.6, -0.3), (-0.3, 0.0)$, and $(0.0, 0.4)$). We mask the region between \logg$=2.0$ and $3.0$ to avoid the RC stars and fit the distribution of the rest RGB stars with the following empirical model,
\begin{equation}\label{eq:rgbmodel1}
N_{\rm RGB}(T_{\rm eff},{\rm {log}}g) = N({\rm log}g)*exp(-\frac{(T_{\rm eff}-T({\rm log}g))^2}{\sigma^2({\rm log}g)})
\end{equation}
where $N({\rm log}g)$ and $\sigma^2({\rm log}g)$ are three smoothing cubic spline functions with different metallicity bins ($\feh\,=(-0.6, -0.3), (-0.3, 0.0)$, and $(0.0, 0.4)$), respectively; and

\begin{equation}\label{eq:rgbmodel2}
T({\rm log}g)=\left\{
\begin{array}{l l}
150.2{{\rm log}g}^3-946.3{{\rm log}g}^2+2417{{\rm log}g}+2188 & \feh\,\in\,(-0.6,-0.3) \\
49.8{{\rm log}g}^3-374.1{{\rm log}g}^2+1435{{\rm log}g}+2632 & \feh\,\in\,(-0.3,0.0) \\
81.65{{\rm log}g}^3-590.7{{\rm log}g}^2+1921{{\rm log}g}+2107 & \feh\,\in\,(0.0,0.4)
\end{array}
\right.
\end{equation}

\begin{figure}
\centering
\includegraphics[scale=0.6]{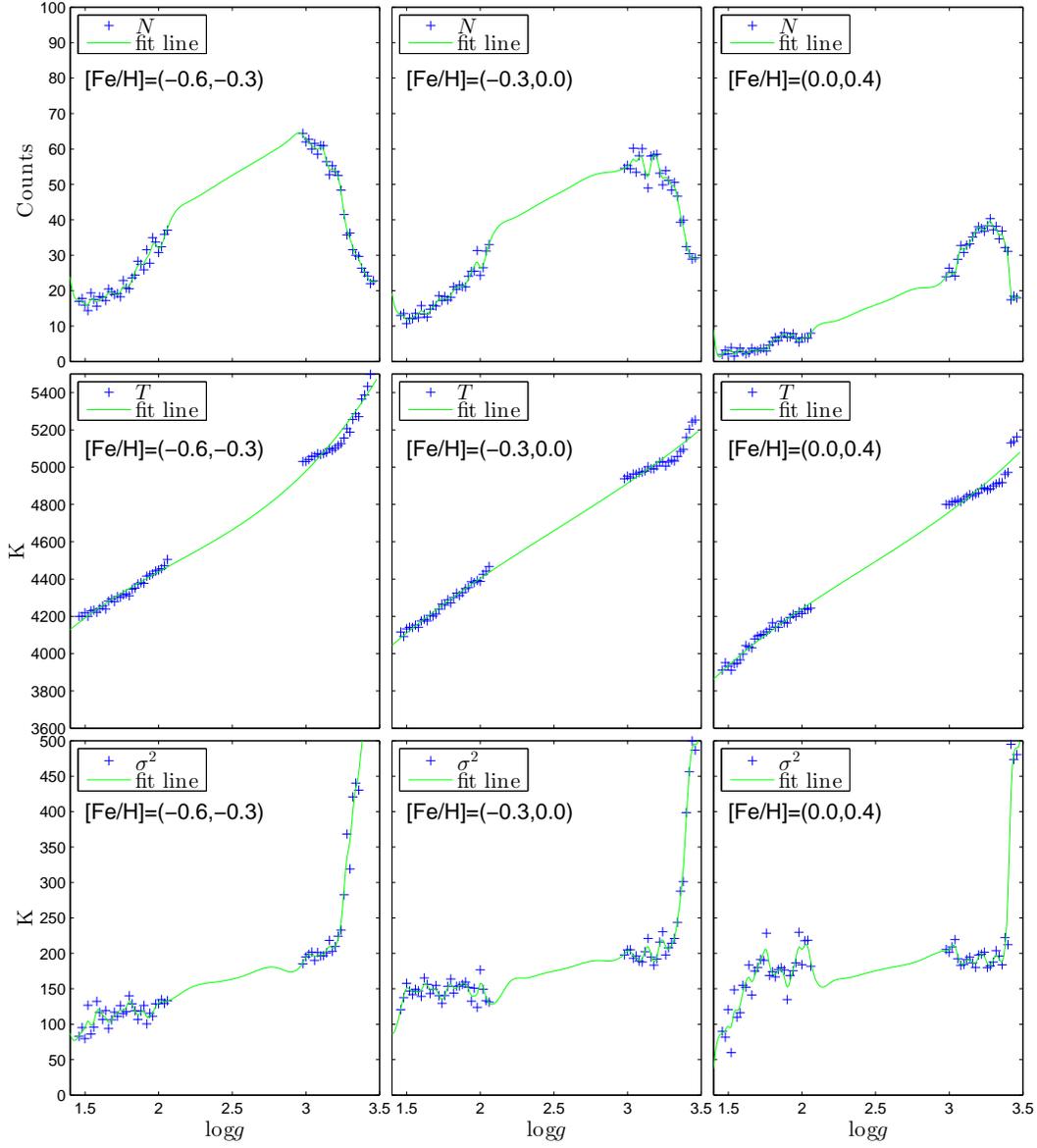}
\caption{Fitting the coefficients in Eq~\ref{eq:rgbmodel1}. The solid lines are the best fit curve, while the cross symbols are measured from RGB stars with \logg $<2$ and $>3$. From left to right, the metallicity bins are $(-0.6, -0.3)$, $(-0.3, 0.0)$, and $(0.0, 0.4)$, respectively. The top panels show the best fit smoothing splines for $N$, the bottom panels show the best fit splines for $\sigma^2$. The rest panels show the best fit polynomials, which coeffecients are appeared in Eq~\ref{eq:rgbmodel2}.}
\label{fig2}
\end{figure}

Fig~\ref{fig2} shows the best fit curves for the terms in Eq~\ref{eq:rgbmodel1}-~\ref{eq:rgbmodel2}. From left to right, the range of metallicity is $(-0.6,-0.3), (-0.3, 0.0)$, and $(0.0, 0.4)$, respectively. $N({\rm log}g)$ and $\sigma^2({\rm log}g)$ are the best fit smoothing spline functions shown in the top panels and bottom panels of Fig~\ref{fig2}, respectively. The rest panels of Fig~\ref{fig2} show the best fit polynomials for the terms in Eq~\ref{eq:rgbmodel2}.

\begin{figure}
\centering
\includegraphics[width=\textwidth]{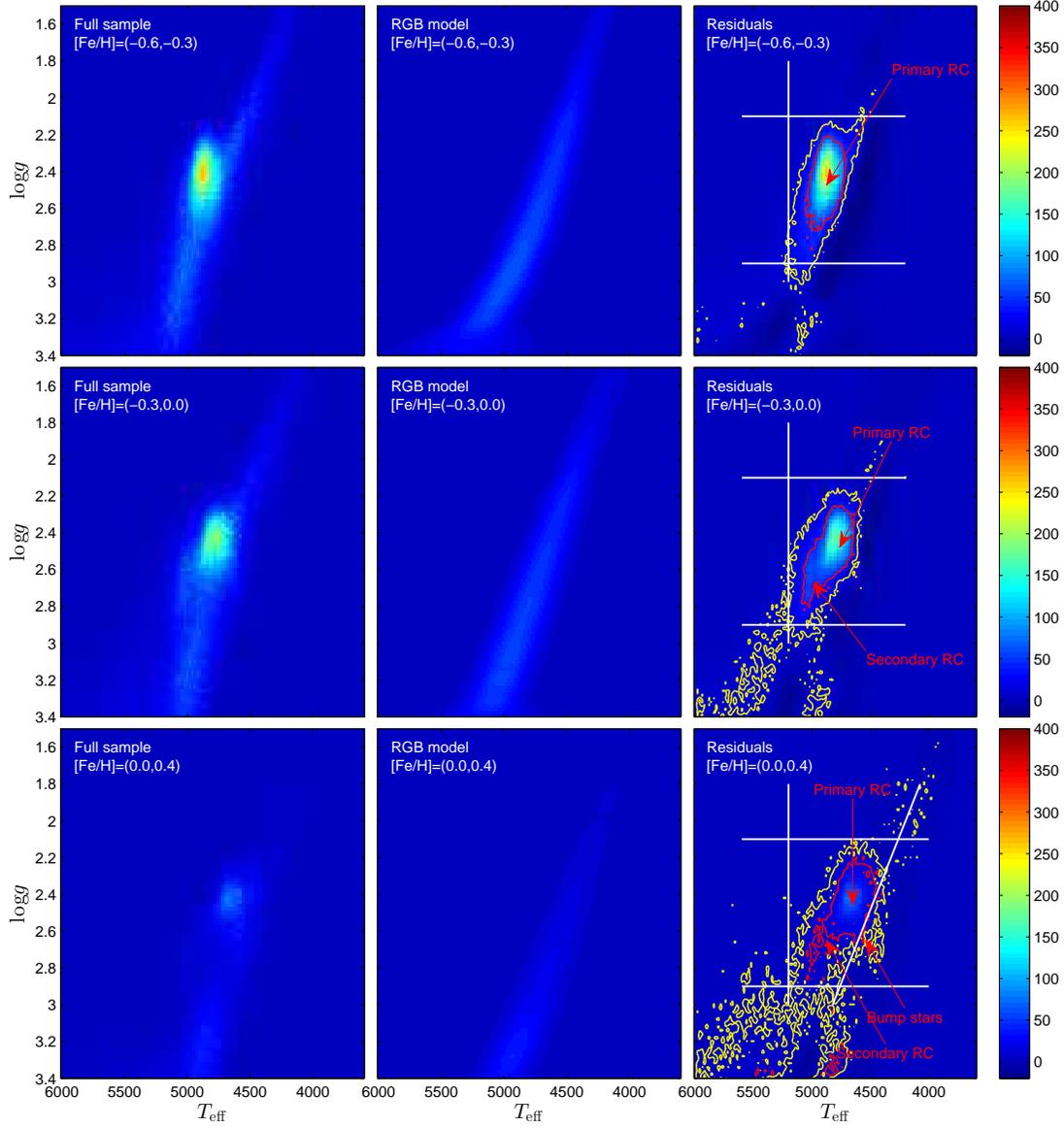}
\caption{From top to bottom, the metallicity bins are $(-0.6, -0.3)$, $(-0.3, 0.0)$, and $(0.0, 0.4)$, respectively. The left panels present the distributions of the full sample of the giant stars with different metallicity, respectively. The middle panels present the best fit distribution models of RGB stars. The right panels present the distributions of the residuals of full sample in the left panels after subtracting the RGB distributions in the middle panels. They are mostly contributed by the RC stars. The red and yellow contours show the 68\% and 95\% completeness of the RC candidates, respectively. The horizontal and vertical white lines give cuts for stars with \logg\,$>2.9$ dex, \logg\,$<2.1$ dex and \teff\,$>5200$\,K, respectively. The leaning white line located at \logg$ =0.0016$\teff$-4.7170$ gives another cut for removing the bump stars located at the bottom-right corner at the 95\% contour in the bottom-right panel. The color codes the stellar count in the bins.}
\label{fig3}
\end{figure}

Providing that the RGB stars are smoothly distributed along \logg, we can \emph{interpolate} the stellar distribution of the RGB stars between \logg$=2.0$ and $3.0$ with the best fit model shown in Eqs~\ref{eq:rgbmodel1}-~\ref{eq:rgbmodel2}. The middle panels of Fig~\ref{fig3} shows the distribution of RGB stars according to our model in \teff\ vs. \logg\ plane for various \feh\ bins in different rows. 

Second, we subtract the smoothing distribution model of RGB stars between \logg$=2$ and $3$ and the residuals are mostly contributed by the RC stars, as shown in the right panels of Fig~\ref{fig3}, in which the contours of 68\% (red) and 95\% (yellow) completeness of the residual distribution are overlapped. Compromise has to be made between the completeness for both primary and secondary RC stars and the fraction of the contamination from the RGB stars. We find that the 68\% completeness contour cannot cover the most region of the secondary RC stars, therefore, we select the 95\% completeness, as the recommended selection criterion of the RC stars. With this empirical distribution of RC stars, other users can freely adjust the criterion of selection to identify different sample of the RC stars to meet their specific requirements. It is noted that there are a few fractional areas far way from the empirical RC star region are also within the 95\% completeness level. They may be artificial features because the data is too sparse in these regions. We then manually exclude three artificial areas with \teff$>5200$\,K, \logg$<2.1$\, dex and \logg$>2.9$\,dex. Moreover, the bottom-right corner of the 95\% completeness level in the bottom-right panel is apparently contributed by the red bump stars rather than the clump stars. Therefore, a fourth cut at \logg$<0.0016$\teff$-4.7170$ is added to remove the contamination from the bump stars. These additional data cuts are shown as white lines in the right panels of Fig~\ref{fig3}. Finally, we find 118,711 RC candidates with the refined 95\% completeness level.

\begin{figure}
\centering
\includegraphics[scale=0.6]{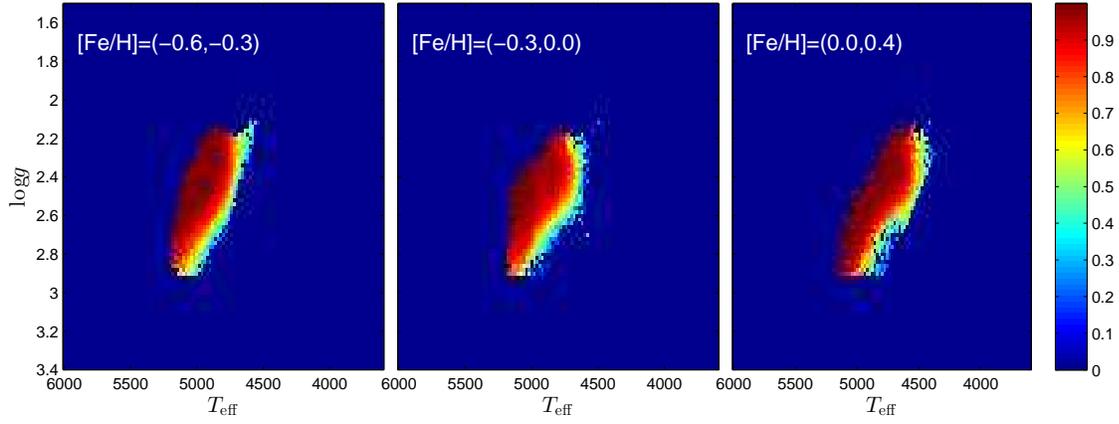}
\caption{The colors show the fractions of true RC stars with different metallicity bins, $(-0.6, -0.3)$, $(-0.3, 0.0)$, and $(0.0, 0.4)$, from left to right panel, respectively, in the ${\rm log}g$ vs. $T_{\rm{eff}}$ plane.}
\label{fig4}
\end{figure}

Although most of the stars located in the refined 95\%-level region are RC stars, some contaminations may also included. Assuming that the residual distribution in \teff\ vs. \logg\ planes of Fig~\ref{fig3} are the distributions of the true RC stars, we can give the percentage of the true RC stars by dividing the residual distributions with the full distributions, which contain both RC and RGB stars, although we cannot identify the individual RC stars. Fig~\ref{fig4} shows the percentage in \teff\ vs. \logg\ planes for metallicity bin $(-0.6, -0.3)$, $(-0.3, 0.0)$, and $(0.0, 0.4)$ from left to right, respectively. It shows that the fraction of the true RC stars are mostly larger than 75\%, even for some regions of the secondary RC stars, in the candidate catalog. The contour level of the 68\% and 95\% completeness, the numbers of the RC candidates under such completeness levels, and the total fraction of the true RC stars are listed in Table~\ref{table1}.

Compared with the method provided by Bovy et al.~(\cite{bovy14}), our method does not depend on the stellar model but only on the specific observations. Moreover, although the number of the secondary RC are less than that of the primary RC, we can still discriminate them from the background RGB stars with a fraction of 75\%$\sim$85\%, as shown in Fig~\ref{fig4}. Therefore, the identification of RC stars in this work is suitable for both primary and secondary RC stars, ensuring that the study of the evolution of the Milky Way can be extended from $<$1\,Gyr to around 10\,Gyr based on this sample.

\section{Distance estimations}\label{sect:dist}
Most of the RC stars are located in the Galactic disk and their apparent magnitudes and color indices are significantly affected by the interstellar extinction. Therefore, the distance and the reddening have to be determined simultaneously. In the next section, we introduce a likelihood method to determine these quantities, and then we apply this method with fixed and various absolute magnitude, respectively, in the following two sections.

\subsection{A likelihood method to estimate distance and reddening}\label{sect:dist1}
The first step to estimate the distance of RC stars is to estimate the reddening from the observed color index. The LAMOST spectra do not have a perfect optical multi-band photometry by now. Its input catalog is a combination of UCAC4 (Zacharias et al.~\cite{zac13}), PanSTARRS1 (Tonry et al.~\cite{ton12}), SDSS (Ahn et al.~\cite{ahn14}), and Xuyi Galactic anti-center survey (Yuan et al.~\cite{yuan15}). Although all of these source catalogs contains $g$, $r$, and $i$ bands, they are still not well calibrated with each other, yet. Therefore, in this stage, we use the 2MASS catalog (Skrutskie et al.~\cite{skr06}) as the input catalog to derive the reddening and distance for RC stars.

The likelihood of $E(J-K)$ for a RC star given the observed $J-K$ and the intrinsic color index of RC stars can be written as:
\begin{equation}\label{eq:ejk}
Pr(E(J-K)|J-K, \sigma_{J-K}, (J-K)_{RC}, \sigma_{RC,J-K})\sim exp(-\frac{(E(J-K)-((J-K)-(J-K)_{RC}))^2}{2({\sigma}^2_{J-K}+{\sigma}^2_{RC,J-K})}),
\end{equation}
where $(J-K)_{RC}$ is the intrinsic color index of RC stars, $\sigma_{RC,J-K}$ the intrinsic color index dispersion of RC stars, and $\sigma_{J-K}$ the measurement error of the observed $J-K$. To convert the reddening in $J-K$ to the extinction in $K$ band, we adopt Indebetouw et al.~(\cite{ind05}) that
\begin{equation}\label{eq:ak}
A_{\rm K} = 0.67E(J-K).
\end{equation}
Then, the likelihood of the distance module, $DM$, for a RC star given the observed $K$ magnitude and the fixed absolute magnitude ${\rm M}_K$ can be written as:
\begin{equation}\label{eq:dmrc}
Pr(DM|K, \sigma_K,A_K, {\rm M}_K, \sigma_{{\rm M}_K})\sim exp(-\frac{(DM-(K-A_{K}-{\rm M}_{K}+5))^2}{2({\sigma}^2_{K}+{\sigma}^2_{{\rm M}_{K}})}),
\end{equation}
where ${\rm M}_K$ is the fixed absolute magnitude in $K$ band for RC stars, $\sigma_{{\rm M}_K}$ the intrinsic dispersion of absolute magnitude in $K$ band for RC stars, and $\sigma_{K}$ is the measurement error of the apparent $K$ magnitude.

\subsection{The fixed absolute magnitude and the intrinsic color index of the RC stars}\label{sect:fixedmk}
Combined with Eqs.~\ref{eq:ejk}-~\ref{eq:dmrc}, the likelihood of both $E(J-K)$ and $DM$ for RC stars can be derived. The last ingredient needed to put in is the absolute magnitude, intrinsic color index, and their intrinsic dispersions for RC stars. Although some literatures provided the absolute magnitude in $K$ band and the intrinsic absolute magnitude in $J-K$ for RC stars (Alves~\cite{alv00}; Groenewegen~\cite{gro08}; Zasowski et al.~\cite{zasowski13} etc.), the intrinsic dispersions for both quantities, which are necessary in our likelihood method, are not self-consistently provided. Therefore, we estimate the absolute magnitude in $K$ band, the intrinsic color index in $J-K$, and their intrinsic dispersions with the \emph{Hipparcos} data. 

\begin{figure}
\begin{center}
\includegraphics[scale=0.8]{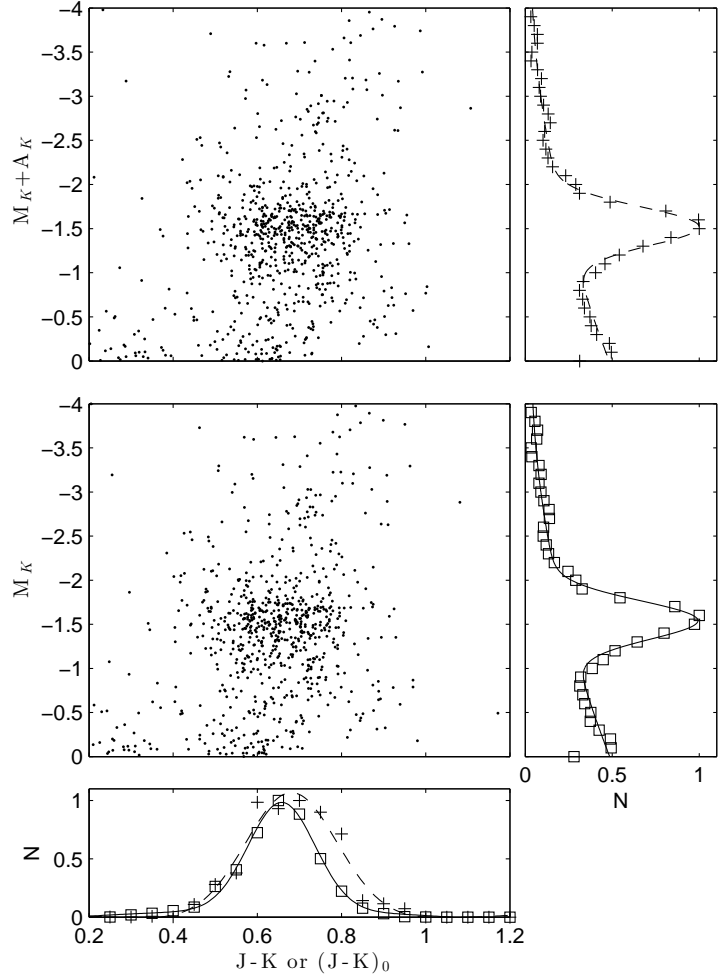}
\caption{The top-left panel shows the reddened $J-K$ vs. ${\rm M}_K+A_K$ diagram for about 900 giant stars selected from \emph{Hipparcos} catalog. The middle-left panel shows the similar plot but with the dereddened color index and absolute magnitude. The bottom panel shows the marginalized distribution of the reddened $J-K$ for the stars (corresponding to the top-left panel) with cross symbols and that of the dereddened color index $(J-K)_0$ (corresponding to the middle-left panel) with the square sybmols. The dashed and solid lines are the best fit model of Eq~(\ref{eq:jkmodelhip}) for the $J-K$ and $(J-K)_0$, respectively. The top-right panel shows the marginalized distribution of ${\rm M}_K+A_K$ with cross symbols. The dashed line stands for the best fit model according to Eq.~(\ref{eq:mkmodelhip}). The middle-right panel shows the marginalized distribution of the dereddened $M_K$ with square symbols, and the best fit model with the solid line.}
\label{fig:hipJKMK}
\end{center}
\end{figure}

\emph{Hipparcos} catalog provides parallaxes for more than 100,000 bright stars, thousands of which located in the region of red clump in HR diagrams. In order to estimate the absolute magnitude of RC stars, we need to correct the reddening as the first step. Bailer-Jones~(\cite{calj11}) estimated the extinction parameters for about 47,000 \emph{Hipparcos} stars, sparsely distributed in the sky. We extend the extinction to all \emph{Hipparcos} stars using spatial interpolation. For a star of interest, we select all stars with reddening parameters derived by Bailer-Jones~(\cite{calj11}) within a 10-degree-radius circle and 20\,pc in distance around it. Then we assign the median reddening value for all selected stars as the reddening value for the star of interest. In order to ensure the accuracy of the absolute magnitude, we select the stars with error of parallax smaller than 20\% and errors of 2MASS photometry smaller than 0.5\,mag. Fig~\ref{fig:hipJKMK} shows the $J-K$ vs. ${\rm M}_K$ diagrams without (top-left panel) and with dereddening (middle-left panel) for about 900 giant stars with ${\rm M}_K<0$\,mag. 

We adopt the empirical model of the distribution in ${\rm M}_K$ from Paczy\'{n}ski \& Stanek~(\cite{pac98}), which has the following form:

\begin{equation}\label{eq:mkmodelhip}
F=a+b({\rm M}_K-c)^2+d{\rm exp}(-{({\rm M}_K-e)^2\over{2f^2}}),
\end{equation}
where $a$, $b$, $c$, $d$, $e$, and $f$ are the free parameters. The quadratic polynomial in Eq~(\ref{eq:mkmodelhip}) models the stellar distribution of the RGB stars and the Gaussian term models the RC stars.

Similarly, we use this model for the marginalized distribution of color index:
\begin{equation}\label{eq:jkmodelhip}
F'=a'+b'(J-K-c')^2+d'{\rm exp}(-{(J-K-e')^2\over{2f'^2}}).
\end{equation}

We fit the models for the absolute magnitude and color index without and with the dereddening, respectively. The top-right panel of Fig~\ref{fig:hipJKMK} shows the best fit for the marginalized reddened absolute magnitude with Eq~(\ref{eq:mkmodelhip}). The middle-right panel shows the best fit model for the marginalized dereddened absolute magnitude. And the bottom panel shows the best fit for the reddened (dashed line) and dereddened (solid line) color index $J-K$, respectively. Table~\ref{tab:RCMKintrinsic} lists the best fit absolute magnitude and intrinsic color index and their dispersions. It shows that the dereddend absolute magnitude and intrinsic color index are brighter and bluer than the reddened values by about 2\%, respectively. In this work we adopt the dereddened ${\rm M}_K$ and $J-K$ as the standard value in the estimation of the distances for RC stars.

\subsection{The \MK\ based on the synthetic isochrones}\label{sect:dist2}
The assumption that the RC stars have a fixed magnitude can only work for the primary RC stars, which are mostly composed of the relatively older RC stars compared with the secondary ones. When we want to trace the evolution of the Galactic disk with RC stars, we cannot only use the primary RC stars and ignore the secondary RC stars. Therefore, we need to improve the approach to estimate the distance of RC stars so that the secondary RC stars are also taken into account.

\begin{figure}
\begin{minipage}[t]{0.45\linewidth}
\centering
\includegraphics[width=\textwidth]{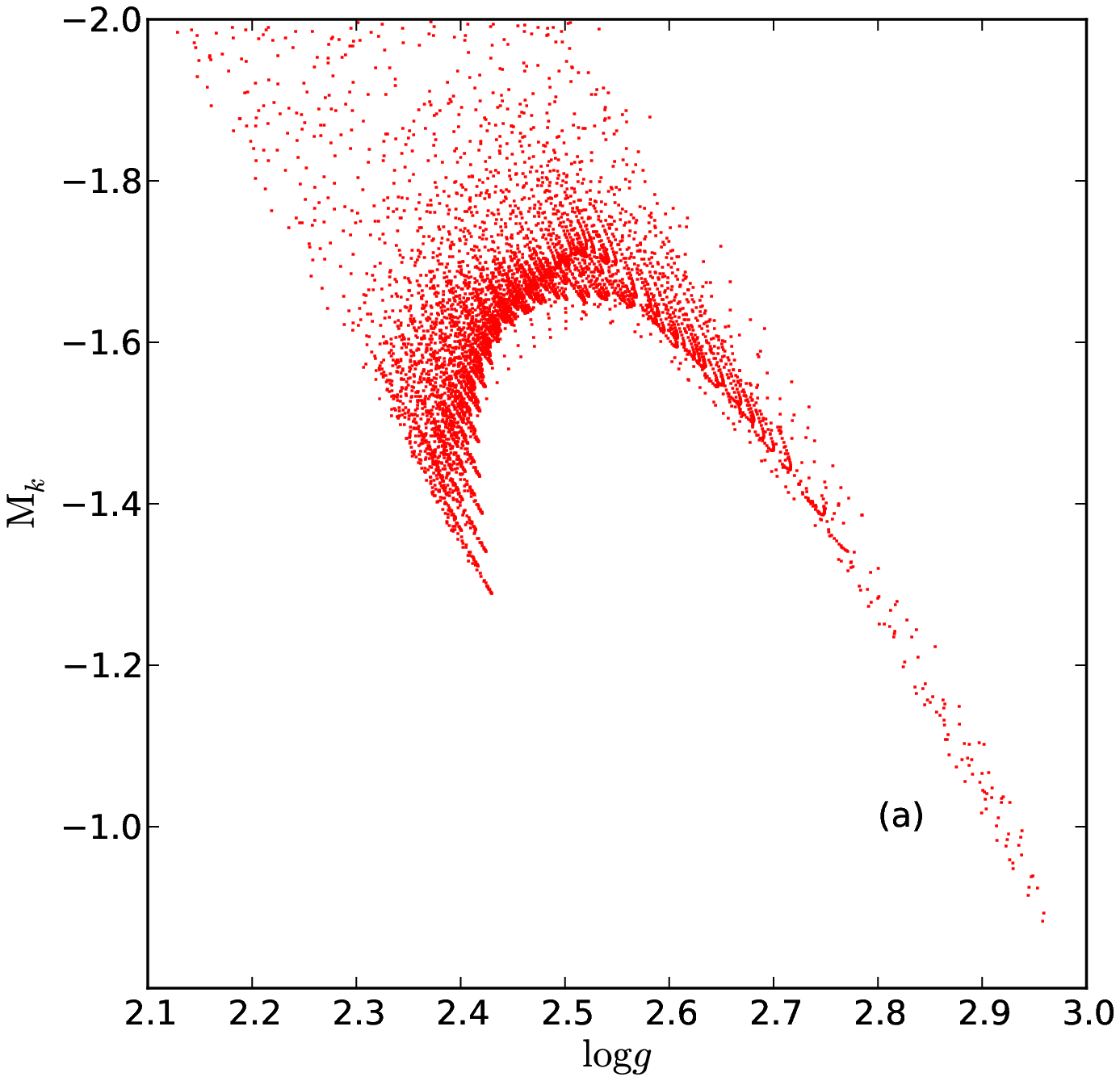}
\end{minipage}
\begin{minipage}[t]{0.45\linewidth}
\centering
\includegraphics[width=\textwidth]{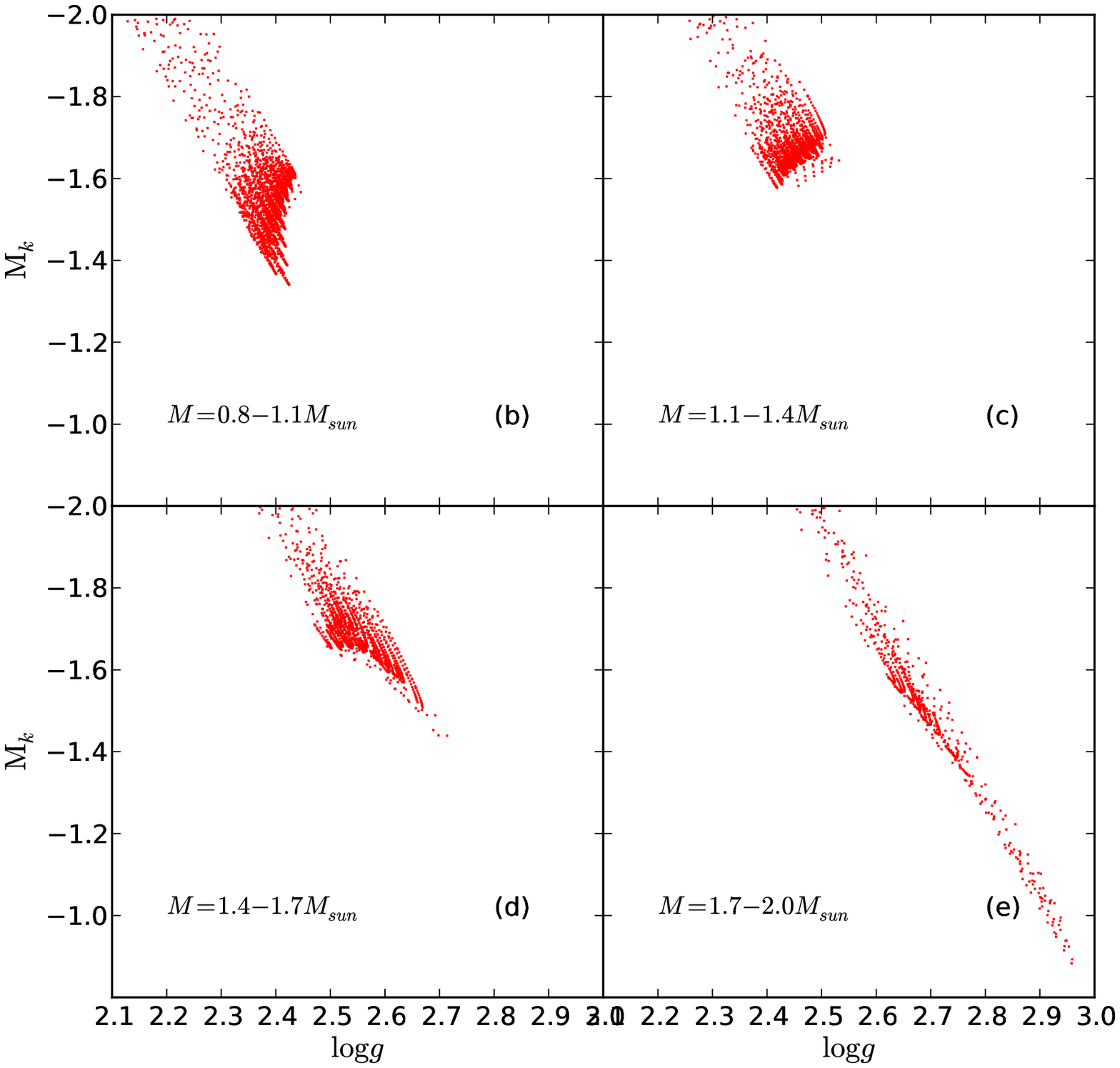}
\end{minipage}
\caption{Panel (a): The synthetic ${\rm log}g$ vs. $ {\rm M}_{K}$ diagram for RC stars from PARSEC stellar evolution track(Bressan et al. ~\cite{bres12}). The range of Age is from 4Myr to 13Gyr with steps of $\Delta({\rm log}t)=0.05$. The ranges of Metallicity \feh\, and  $M_{\rm ini}$ are $-0.6\sim0.3$\, ($Z_{\sun}$=0.0152) and $0.8\sim2M_{\sun}$ respectively. Panel (b)-(e): The ${\rm log}g$-${\rm M}_{K}$ relation for RC stars with $M_{\rm ini}$ between $0.8\sim1.1M_{\sun}$, $1.1\sim1.4M_{\sun}$, $1.4\sim1.7M_{\sun}$, and $1.7\sim2.0M_{\sun}$, respectively.}
\label{fig5}
\end{figure}

\begin{figure}
\centering
\includegraphics[width=0.85\textwidth]{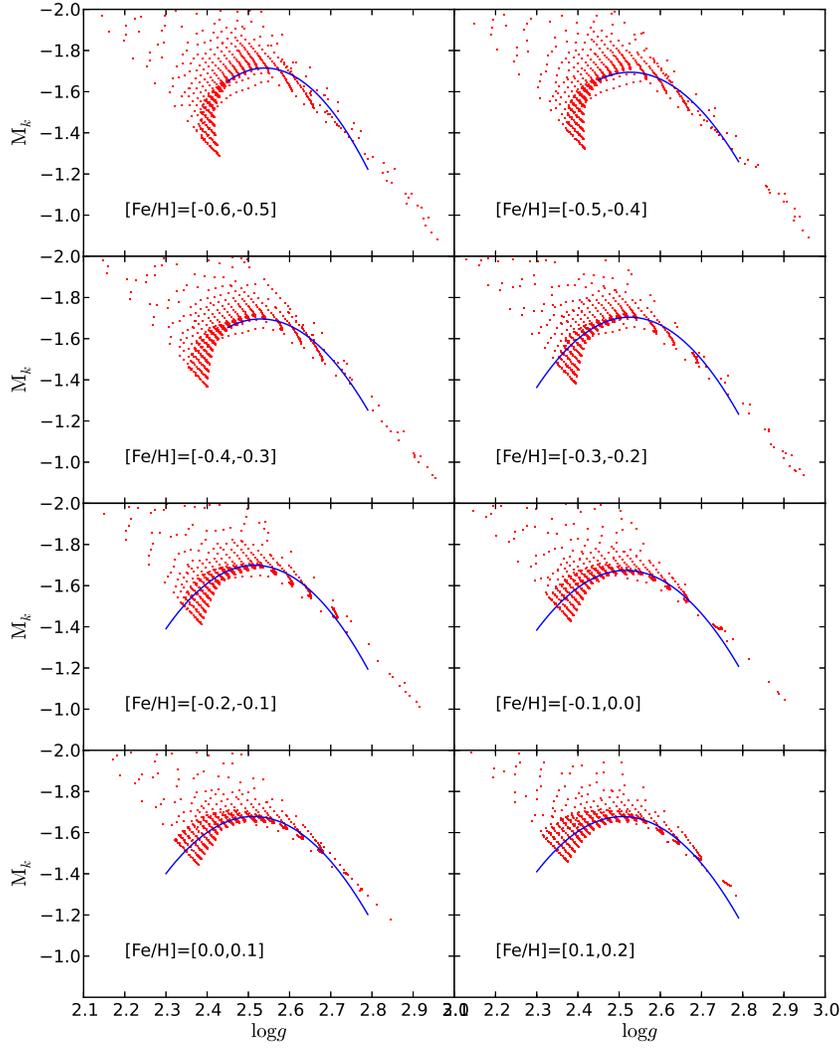}
\caption{For RC stars with ${\rm log}g>2.45$, the ${\rm M}_{K}$ is modeled as quadratic polynomials of ${\rm log}g$ in each \feh\, bin. The red dots are the synthetic data and the blue lines are the best fit quadratic polynomials.}
\label{fig6}
\end{figure}

\begin{figure}
\centering
\includegraphics[width=0.75\textwidth]{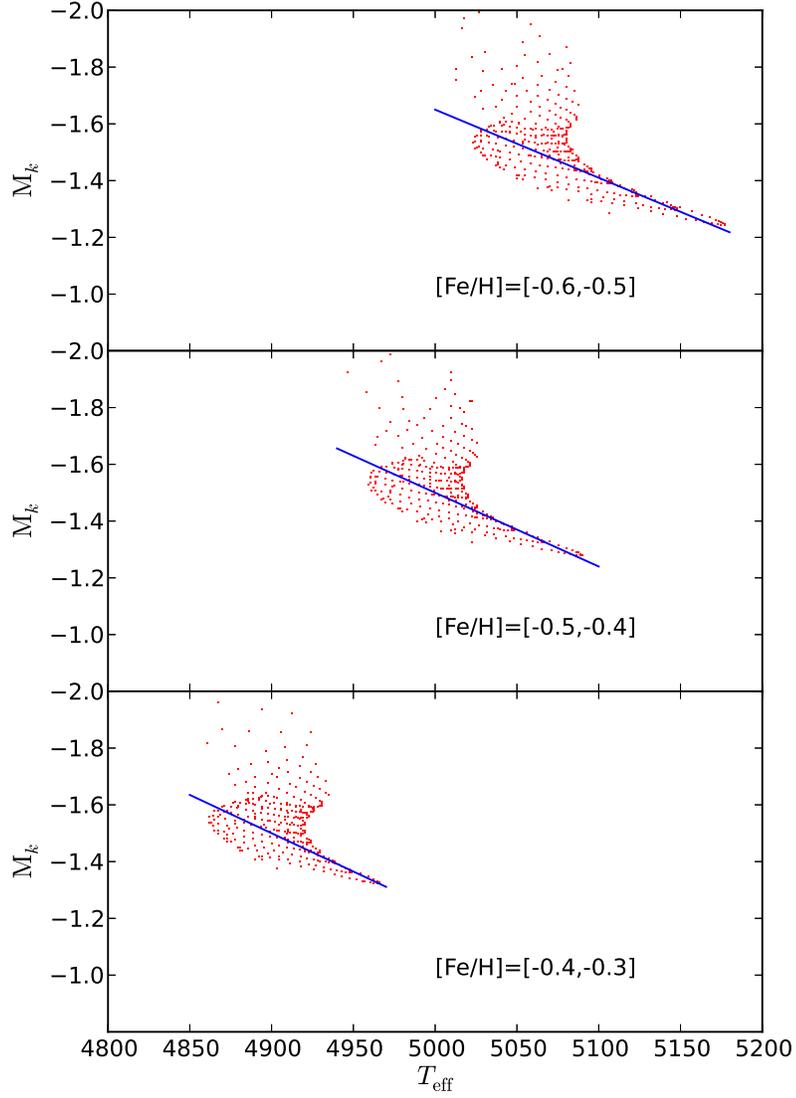}
\caption{For RC stars with ${\rm log}g<2.45$ and \feh$<-0.3$, the ${\rm M}_{K}$ is modeled linearly as a function of \teff\, in each \feh\, bin. The red dots are the synthetic data and the blue lines are the best fit lines.}
\label{fig7}
\end{figure}

We turn to use a special isochrone fitting process to estimate the absolute magnitude for all kinds of RC stars. After quickly reviewing the isochrones, we realize that the absolute magnitude of RC stars is a function of \logg, \feh, and \teff. Panel (a) of Fig~\ref{fig5} shows the synthetic RC stars from PARSEC library (Bressan et al.~\cite{bres12}) in \logg\ vs. ${\rm M}_K$ plane. It shows that ${\rm M}_K$ is tightly dependent of \logg. Further separations of the data into different initial stellar masses are shown in the right panels. We find that the RC stars with stellar masses of $0.8\sim1.1$\,$M_{\sun}$ (panel (b) in Fig~\ref{fig5}) mostly concentrate within ${\rm M}_K=-1.4\sim-1.6$\,mag, while the ${\rm M}_K$ for the RC stars with initial mass of $1.1\sim1.4$\,$M_{\sun}$ moves up to $-1.6\sim-1.8$\,mag (panel (c)). Then the ${\rm M}_K$ for the RC stars with initial mass at $1.4\sim1.7$\,$M_{\sun}$ move back to the range of $-1.5\sim-1.7$\,mag (panel (d)). For the RC stars with $1.7\sim2.0$\,$M_{\sun}$, ${\rm M}_K$ intensively extends from $\sim-1$ to more than $-2$\,mag (panel (e)). To further investigate how the ${\rm M}_K$ varies, we separate the synthetic stars into two groups at \feh$=-0.3$. For the RC stars with \feh$>-0.3$\,dex, the relation between ${\rm M}_K$ and \logg\ can be empirically modeled with a quadratic polynomial (see the related panels in Fig~\ref{fig6}):
\begin{equation}
{\rm M}_{K} = P_{1}{{\rm log}g}^2 + P_{2}{\rm log}g + P_{3}
\end{equation}
The best fit coeffecients $P_i$ ($i=1,2,3$) are listed in Table~\ref{table2}.
For the stars with \feh$<-0.3$\,dex, when \logg$<2.45$, the $M_K$ is no longer a function of \logg\ (see the related panels in Fig~\ref{fig6}). Fig~\ref{fig7} shows that for these stars, the absolute magnitude is roughly a linear function of the effective temperature. Then we have the following more complicated relation: 
\begin{equation}
{\rm M}_K=\left\{
\begin{array}{l l}
P_{1}T_{\rm eff} + P_{2} &  {\rm log}g < 2.45 \\
P_{1}{{\rm log}g}^2 + P_{2}{\rm log}g + P_{3} &  {\rm log}g>2.45
\end{array}
\right.
\end{equation}
Table~\ref{table3} shows the best fit coefficients of $P_i$ ($i=1,2,3$). It is noted that for both groups of metallicity, the synthetic data (red dots) showing in Figs~\ref{fig6} and~\ref{fig7} are not exactly located on a narrow line, but spread out with various dispersions. We then measure the dispersions of the residuals of the ${\rm M}_K$ for the synthetic RC stars with respect to the best fit models at different bins of metallicity and show them in the column of $\sigma_{{\rm M}_K}$ in Tables~\ref{table2} and~\ref{table3}.

Then, for each observed RC star, we firstly derive the ${\rm M}_K$ and $\sigma_{{\rm M}_K}$ from its \teff, \logg, and \feh, and bring them back to Eq~\ref{eq:dmrc} to derive the likelihood of the distance module. Because the intrinsic color index of the primary and secondary RC stars are quite similar, we adopt this value derived from section~\ref{sect:fixedmk}.

\subsection{Performance assessment}\label{sect:perform}
Before applying the improved distance estimation to the LAMOST data, we assess the performance of the improved ${\rm M}_K$ model demonstrated in section~\ref{sect:dist2}. 

\begin{figure}
\centering
\includegraphics[width=\textwidth]{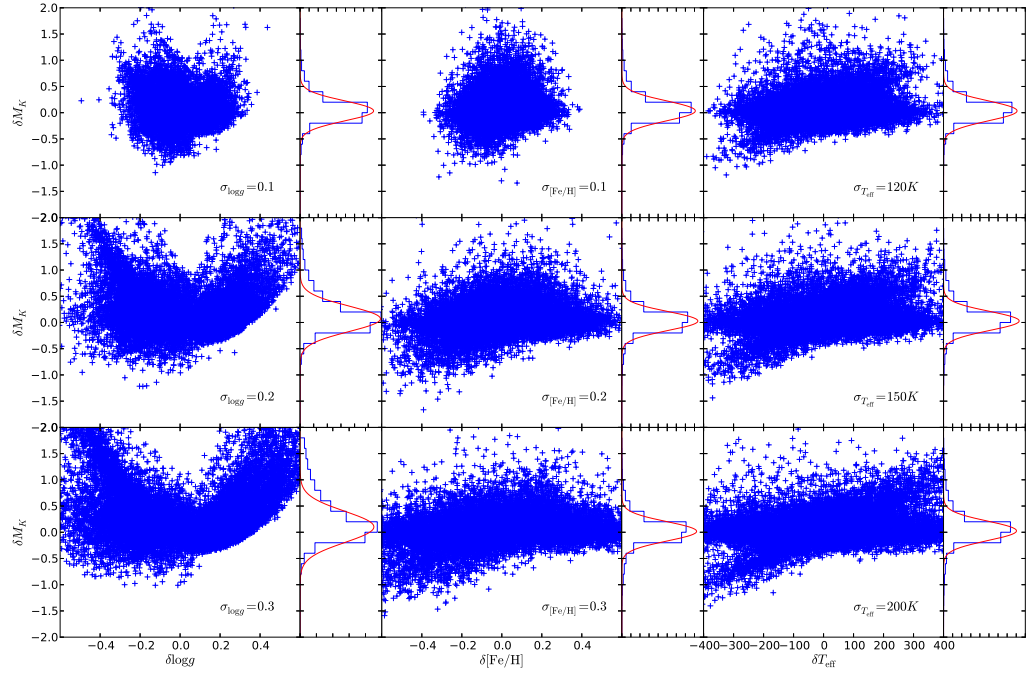}
\caption{The scatter plot of the residuals of derived ${\rm M}_{K}$, denoted by $\delta {\rm M}_{K}$, for 20,000 simulate data with various uncertainties in \logg\, (left panels), \feh\, (middle panels), and \teff\, (right panels).}
\label{fig8}
\end{figure}

We arbitrarily select 1000 points from the synthetic dataset and add random Gaussian errors to the true values of  \teff, \logg, and \feh. For each synthetic RC star, we create 20 mock stars with different random errors. In total, we create 20,000 mock stars with errors. Then, we derive the absolute magnitude for the mock data based on the method mentioned in Section~\ref{sect:dist2}. We totally create 9 sets of mock dataset with various measurement errors of \logg, \teff, and \feh. In the first three mock datasets, we create the errors with $\sigma_{{\rm log}g}=0.1$, 0.2, and 0.3\,dex, respectively, $\sigma_{T_{\rm eff}}=120$\,K, and $\sigma_{\rm[Fe/H]}=0.1$\,dex. The second three mock datasets are created with random errors of $\sigma_{T_{\rm eff}}=120$, 150, and 200\,K, respectively, $\sigma_{{\rm log}g}=0.1$\,dex, and $\sigma_{\rm[Fe/H]}=0.1$\,dex. The last three mock datasets have $\sigma_{\rm[Fe/H]}=0.1$, 0.2, and 0.3\,dex, respectively, with $\sigma_{{\rm log}g}=0.1$\,dex, and $\sigma_{T_{\rm eff}}=120$\,K. We compare the derived absolute magnitudes with the true values in the 9 mock samples. The residuals of the derived absolute magnitudes as functions of the errors of the stellar parameters are shown in Fig~\ref{fig8}.

\begin{figure}
\centering
\includegraphics[width=\textwidth]{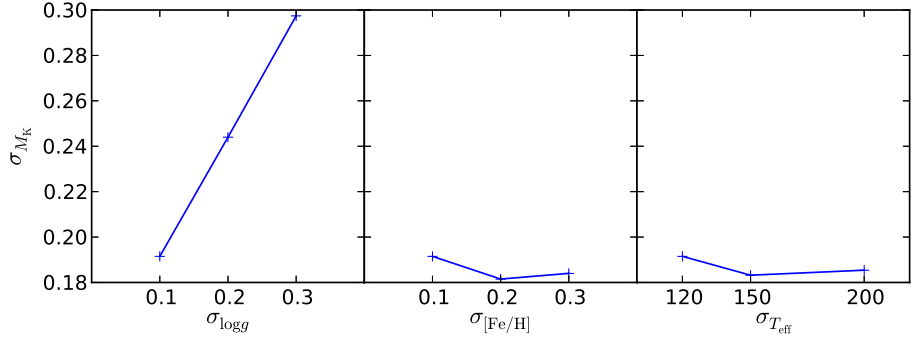}
\caption{The relationship between the standard deviation of the residual ${\rm M}_{K}$, denoted by $\sigma_{{\rm M}_{K}}$, and the uncertainty of \logg\, ($\sigma_{{\rm {log}}g}$), \feh\, ($\sigma_{{\rm [Fe/H]}}$), and \teff\, ($\sigma_{T_{\rm eff}}$).}
\label{fig9}
\end{figure}

The left panels of Fig~\ref{fig8} show the residuals of ${\rm M}_K$, denoted as $\delta {\rm M}_K$, as a function of the errors of \logg\ at $\sigma_{{\rm log}g}=0.1$, 0.2, 0.3\,dex from top to bottom, respectively. It shows that the larger the uncertainty of \logg, the larger the errors in the absolute magnitude. When the random errors of \logg\ are larger, the derived ${\rm M}_K$ seems more overestimated. However, slightly increasing the uncertainties of \feh\ (the middle panels) and \teff\ (the right panels) would not significantly add the uncertainties of the ${\rm M}_K$ estimates. Fig~\ref{fig9} shows the standared deviation of $\delta {\rm M}_K$, in terms of the $\sigma$ of the best fit Gaussian with the histogram of $\delta {\rm M}_K$, as functions of the uncertainties of \logg\ (the left panel), \feh\ (the middle panel), and \teff\ (the right panel), respectively. Again, this figure shows that the accuracy of the ${\rm M}_K$ estimates relies mostly on the accuracy of \logg, rather than that of \feh\ and \teff. Therefore, the accurate \logg\ calibrated with the asteroseismic \logg\ from Liu et al.~(\cite{liu15}) is very important in the high accuracy distance estimation.

Figure~\ref{fig9} also shows that with the typical uncertainty of 0.1\,dex in \logg, the uncertainty of the derived absolute magnitude is better than 0.2\,mag, corresponding to $\sim$10\% in distance. 

\section{Discussions}\label{sect:disc}
\subsection{Comparison between two absolute magnitude models}
\begin{figure}
\centering
\includegraphics[width=\textwidth]{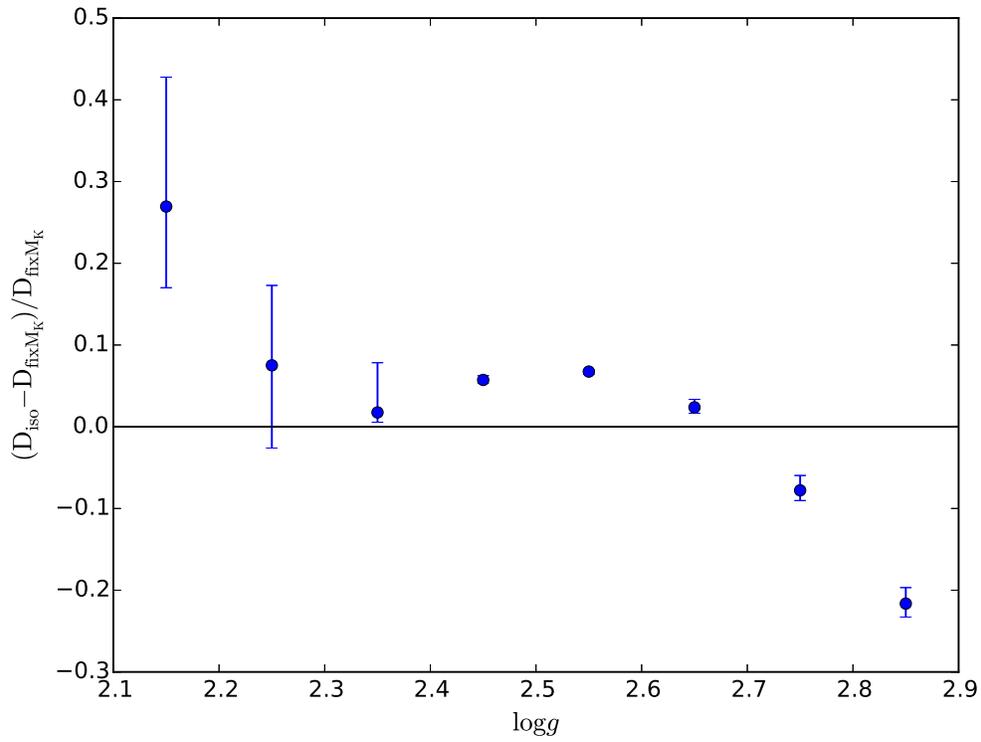}
\caption{The comparison between the the isochrone-based ($\rm D_{iso}$) and fixed absolute magnitude-based ($\rm D_{{fixM}_{\textit K}}$) distance estimates at various \logg\,.}
\label{fig10}
\end{figure}

In section~\ref{sect:dist}, we discuss two approaches to estimate the absolute magnitude for RC stars. It is worthy to directly compare the distance estimates based on the two different methods. Fig~\ref{fig10} shows the difference of the distance estimates for the LAMOST RC stars between the fixed absolute magnitude-based and the isochrone-based method as a function of \logg. It is seen that the fixed magnitude-based method tends to underestimate the RC stars with smaller \logg\ and significantly overestimate those with larger \logg. The overestimation in large \logg\, data is because that the fixed absolute magnitude are dominated by the primary RC stars and hence may not be suitable for the secondary RC stars. The underestimation when \logg\,$<2.7$\,dex is likely because the slightly inconsistency between the ${\rm M}_K$ used in the synthetic isochrones and the one derived in section~\ref{sect:fixedmk}. In Fig~\ref{fig10}, we find that the systematic bias can shift by more than 20\% in large \logg\, given a fixed absolute magnitude. When \logg\,$<2.3$\,dex, the isochrone-based method may not give reliable ${\rm M}_K$ estimates for RC stars since this is very close to the boundary of the isochrone data (see Fig~\ref{fig7}).Therefore the errors increase in this region, as shown in Fig~\ref{fig10}.

\subsection{External uncertainty of the distance estimation}
In Section~\ref{sect:perform}, we use the mock data created from the synthetic data to test the performance of the distance estimation based on the isochrones. This, however, can only give the internal error but not the external one. We then cross-identify the RC stars from the LAMOST data with \emph{Hipparcos} catalog. Unfortunately, we only obtain less than 10 common RC stars with parallax error less than 50\%. Because most of these RC stars suffers larger uncertainty in parallax, they cannot be treated as the standard stars to investigate the external error of the isochrone-based distance estimates. To solve this issue, we need to wait the coming \emph{Gaia} data (Bailer-Jones~\cite{bai09}), which will release the first catalog in 2016.
  
\section{Conclusions}\label{sect:con}
In this work, we set up an empirical model in \teff\ vs. \logg\ plane to identify the RC stars in the LAMOST DR2 data. The employed approach identifies not only the primary, but also the secondary RC stars. This will be very helpful for the study of the evolution of the Milky Way, because the range of the RC stars can be extended from $<$1\,Gyr (contributed by the secondary RC stars) to 10\,Gyr (contributed by the primary RC stars). Finally, we identify 118,711 RC stars from the LAMOST DR2 data with the 95\% completeness level. The sample of the selected RC stars may be contaminated by the RGB stars with the fraction of about 20\%.

After the identification of the RC stars, we develop two different approaches to estimate their distances as well as the interstellar extinction. The first one is based on the fixed absolute magnitude and intrinsic color index. The accuracy of the fixed absolute magnitude-based approach relies on the accuracy of the determination of the absolute magnitude and the dispersion of the absolute magnitude. Consequently, we revisit the absolute magnitude and its dispersion for RC stars in $K$ band. We adopt the similar empirical model introduced by Paczy\'{n}ski \& Stanek~(\cite{pac98}), but take into account the interstellar extinction for the \emph{Hipparcos} RC stars. Although the extinction is small, it leads to the underestimation of the absolute magnitude by about 2\%.

Although this method is sufficiently accurate for most of the primary RC stars, it is not suitable for the secondary RC stars, which significantly vary in \logg\ and hence also in the absolute magnitude. We then develop the second approach considering both types of RC stars based on the isochrones. With the empirical model, we associate the absolute magnitude of a RC star with its \feh, \logg, and \teff. The more delicate model can reduce the uncertainty of the distance estimates to 10\% for almost all types of RC stars given the error of \logg\ at around 0.1\,dex.

This RC sample well samples the Galactic disk, particularly in the outer disk, allowing us to map the structure, kinematics, and evolution of the Galactic disk in future works. 

\begin{acknowledgements}
This work is supported by the Strategic Priority Research Program ``The Emergence of Cosmological Structures" of the Chinese Academy of Sciences, Grant No. XDB09000000 and the National Key Basic Research Program of China 2014CB845700. CL acknowledges the National Science Foundation of China (NSFC) under grants 11373032, 11333003 and U1231119. Guoshoujing Telescope (the Large Sky Area Multi-Object Fiber Spectroscopic Telescope LAMOST) is a National Major Scientific Project built by the Chinese Academy of Sciences. Funding for the project has been provided by the National Development and Reform Commission. LAMOST is operated and managed by the National Astronomical Observatories, Chinese Academy of Sciences.
\end{acknowledgements}


\clearpage

\begin{table}
\centering
\caption{The location of the RC candidates in the right panel in Fig~\ref{fig3}}
\label{table1}
\begin{tabular}{c|c|c|c|c}
\hline\hline
[Fe/H] \ & \ Completeness level & \ Contour level & \ $N_{\rm RC}$  & \ Total ratio of RC stars \\
\hline
(-0.6,-0.3) & 68\%  & 51.39 & 36099 & 89.68\%  \\
 & 95\%  & 15.46 & 54890 & 80.47\%  \\
\hline
(-0.3,0.0) & 68\%  & 42.26 & 31553 & 83.32\%  \\
 & 95\%  & 10.46 & 48064 & 76.29\%  \\
\hline
(0.0,0.4) & 68\%  & 12.12 & 11291 & 84.87\%  \\
 & 95\%  & 3.01 & 15757 & 81.13\%  \\
\hline\hline
\end{tabular}
\tablecomments{0.8\textwidth}{\\Contour level: the contour level in Fig~\ref{fig3} corresponding to 68\% or 95\% completeness.\\
$N_{\rm RC}$: number of RC candidates enclosed in the contour. (with the addition cut around the edges, see white lines in the right panels of Fig~\ref{fig3})\\
Ratio of RC stars: the total fraction of RC stars to the all metal-rich giant stars within the contour level.}
\end{table}

\begin{table}
\centering
\caption{The derived absolute magnitude and intrinsic color index of RC stars.}\label{tab:RCMKintrinsic}
\begin{tabular}{l|c|c|c|c}
\hline
\hline
& ${\rm M}_K$ & $\sigma_{{\rm M}_K}$  & $J-K$ & $\sigma_{J-K}$\\
 & mag & mag & mag & mag \\
\hline
reddened & $-1.529\pm0.003$ & $0.075\pm0.003$ & $0.681\pm0.003$ & $0.104\pm0.005$\\
dereddened & $-1.549\pm0.003$ & $0.076\pm0.003$ & $0.658\pm0.001$ & $0.072\pm0.001$\\
\hline\hline
\end{tabular}
\end{table}

\begin{table}
\centering
\caption{The coefficients at different \feh\ bins for the ${\rm M}_K$ model of metal-rich RC stars}
\label{table2}
\begin{tabular}{c|cccc}
\hline\hline
\     &\multicolumn{4}{|c}{${\rm M}_{K}=P_{1}{\rm log}^2g+P_{2}{\rm log}g+P_{3}$} \\
\hline
\ [Fe/H] & \ $P_{1}$   & \ $P_{2}$   & \ $P_{3}$  & \ $\sigma_{{\rm M}_{K}}$ \\
\hline
(-0.3,-0.2) & 6.72 & -33.94 & 41.15 &0.12 \\
(-0.2,-0.1) & 6.67 & -33.55 & 40.49 &0.10 \\
(-0.1,0.0) & 6.21 & -31.25 & 37.64 &0.11 \\
(0.0,0.1) & 6.17 & -31.00 & 37.26 &0.10\\
(0.1,0.2) & 6.19 & -31.05 & 37.26 &0.09\\
\hline\hline
\end{tabular}
\end{table}

\begin{table}
\centering
\caption{The coefficients at different \feh\ bins for the ${\rm M}_K$ model of metal-poor RC stars}
\label{table3}
\begin{tabular}{c|ccc|cccc}
\hline\hline
\     &  \multicolumn{3}{|c|}{${\rm log}g<2.45$}    & \multicolumn{4}{c}{${\rm log}g>2.45$} \\
\     &  \multicolumn{3}{|c|}{${\rm M}_{K}=P_{1}T_{\rm eff}+P_{2}$}   &\multicolumn{4}{c}{${\rm M}_{K}=P_{1}{\rm log}^2g+P_{2}{\rm log}g+P_{3}$} \\
\hline
\ [Fe/H] & \ $P_{1}$  & \ $P_{2}$   & \ $\sigma_{{\rm M}_{K}}$ & \ $P_{1}$   & \ $P_{2}$   & \ $P_{3}$  & \ $\sigma_{{\rm M}_{K}}$ \\
\hline
(-0.6,-0.5) & 0.0024 & -13.65 & 0.15 &7.87 & -39.98 & 49.06 & 0.09 \\
(-0.5,-0.4) & 0.0026 & -14.50 & 0.14 &6.28 & -31.74 & 38.41 & 0.08 \\
(-0.4,-0.3) & 0.0027 & -14.73 & 0.14 &6.54 & -33.09 & 40.16 & 0.08 \\
\hline\hline
\end{tabular}
\end{table}

\clearpage

\clearpage

\begin{thebibliography}{15}
\bibitem [2014]{ahn14}Ahn, C. P., Alexandroff, R., Allende P., et al. 2014, \apjs, 211, 16
\bibitem [2000]{alv00}Alves, D. R., 2000, \apj, 539, 732
\bibitem [2009]{bai09}Bailer-Jones, Coryn A. L., 2009, IAU Symposium, 254, 475
\bibitem [2012]{bres12}Bressan, A., Marigo, P., Girardi, L., et al. 2012, \mnras, 427, 127
\bibitem [2011]{calj11}Bailer-Jones, C. A. L., 2011, \mnras, 411, 435
\bibitem [2014]{bovy14}Bovy, J., Nidever, D. L., Rix, H-W., et al. 2014, \apj, 790, 21
\bibitem [2007]{cab07}Cabrera-Lavers, A., Hammersley, P. L., Gonz\'{a}lez-Fern\'{a}ndez. C., et al. 2007, \aap, 465, 825
\bibitem [1997]{cassisi97} Cassisi, S., \& Salaris, M.\ 1997, \mnras, 285, 593
\bibitem [2012]{cui12}Cui, X.-Q., Zhao, Y.-H., Chu, Y.-Q., et al., 2012, RAA, 12, 1197 
\bibitem [2014]{deng12}Deng, L.-C., Newberg, H. J., Liu, C., et al. 2012, RAA, 12, 735
\bibitem [1998]{gira98}Girardi, L., Groenewegen, M. A. T., Weiss, A., et al. 1998, \mnras, 301, 149
\bibitem [1999]{girardi99} Girardi, L.\ 1999, \mnras, 308, 818 
\bibitem [2000]{gira00}Girardi, L.,\ 2000, IAU Joint Discussion, 13, 10
\bibitem [2008]{gro08}Groenewegen, M.,A.T. 2008, \aap,  488, 935
\bibitem [2005]{ind05}Indebetouw, R., Mathis, J. S., Babler, B. L., et al. 2005, \apj, 619, 931
\bibitem [2012]{liu12}Liu, C., Xue, X.-X., Fang, M. et al. 2012, \apjl, 753, 24
\bibitem [2014]{liu14a}Liu, X.-W., Yuan, H.-B., Huo, Z.-Y., et al., 2014, Procceding of IAUS 298, Felzing, S., Zhao, G., \& Walton, N. A. Eds., Cambridge University Press, 310 (arXiv: 1306.5376)
\bibitem [2015]{liu15}Liu, C., Fang, M., Wu, Y., et al. 2015, \apj in press, arXiv:1411.0235
\bibitem [2002]{Lop02}L\'opez-Corredoira, M., Cabrera-Lavers, A., Garz\'{o}n, F., et al. 2002, \aap, 394,883
\bibitem [2013]{nataf13} Nataf, D.~M., Gould, A., Fouqu{\'e}, P., et al.\ 2013, \apj, 769, 88
\bibitem [1998]{pac98}Paczy\'{n}ski, B. \& Stanek, K., 1998, \apjl, 494, 219
\bibitem [1997]{perryman97} Perryman, M.~A.~C., Lindegren, L., Kovalevsky, J., et al.\ 1997, \aap, 323, L49 
\bibitem [2002]{sal02}Salaris, M. \& Girardi, L., 2002, \mnras, 337, 332
\bibitem [2006]{skr06}Skrutskie, M. F., Cutri, R. M., Stiening, R., et al. 2006, \apj, 131, 1163
\bibitem [1996]{sta96}Stanek, K., 1996, \apjl, 460, 37
\bibitem [2013]{stello13}Stello, D., Huber, D., Bedding, T.~R., et al. 2013, \apjl, 765, LL41
\bibitem [2012]{ton12}Tonry, J. L., Stubbs, C. W., Lykke, K. R., et al. 2012, \apj, 750, 14
\bibitem [2011a]{wu11a}Wu, Y., Singh, H. P., Prugniel, P., \ 2011a, \aap, 525, 71
\bibitem [2011b]{wu11b}Wu, Y., Luo, A., Li, H., et al.\ 2011b, RAA, 11, 924
\bibitem [2014]{wu14}Wu, Y., Luo, A., Du, B., et al. 2014, arXiv:1407.1980
\bibitem [2012]{yao12}Yao, S., Liu, C., Zhang H.-T., et al. 2012, RAA, 12, 772
\bibitem [2015]{yuan15}Yuan H.-B., Liu X.-W., Huo, Z.-Y., et al. 2015, \mnras, 448, 855
\bibitem [2013]{zac13}Zacharias, N., Finch, C. T., Girard, T. M., et al. 2013, \apj, 145, 14
\bibitem [2013]{zasowski13}Zasowski, G., Johson, J. A., Frinchaboy, P. M., et al, 2013, \aj, 146, 81
\bibitem [2012]{zhao12}Zhao, G., Zhao, Y.-H., Chu Y.-H., et al. 2012, RAA, 12, 723

\end{thebibliography}
\end{document}